\documentclass[a4paper,twoside,french]{article}
\usepackage{rfia2000}
\usepackage[T1]{fontenc}
\usepackage{babel}
\usepackage{times}
\usepackage{graphicx}
\usepackage{hyperref}

\usepackage{xspace}
\newcommand{\scxp}[1]{\protect{\textsc{#1}}\xspace}
\newcommand{\coron}{\scxp{Coron}}

\begin{document}
%%%%%Pas de date
\date{}
%%%%% Titre gras 14 points
\title{\Large\bf \coron : Plate-forme d'Extraction de Connaissances dans les Bases de Donn\'{e}es
       }
%%%%% Si auteur unique
%\author{L. Auteur \\
%%  Son institut \\
%%  Son addresse \\
%%  Son email}
%%%% pour deux auteurs
%\author{\begin{tabular}[t]{c@{\extracolsep{8em}}c}
%%%% pour trois auteurs
%%%%\author{\begin{tabular}[t]{c@{\extracolsep{6em}}c@{\extracolsep{6em}}c}
%%%% pour quatre auteurs
\author{\begin{tabular}[t]{c@{\extracolsep{1em}}c@{\extracolsep{1em}}c@{\extracolsep{1em}}c@{\extracolsep{1em}}c}
%%%%pour plus d\'ebrouillez-vous !
Baptiste Ducatel$^1$ & Mehdi Kaytoue$^{1}$ & Florent Marcuola$^{1}$ & Amedeo Napoli$^{1}$ & Laszlo Szathmary$^{2}$ \\
\end{tabular}
{} \\
~\\
  $^{1}$ Laboratoire Lorrain de Recherche en Informatique et ses Applications (LORIA)\\
  Campus Scientifique -- BP 239 -- 54506 Vand{\oe}uvre-l\`{e}s-Nancy Cedex (France) \\
  $^{2}$  Département d'Informatique -- Université du Québec à Montréal  (UQAM) \\
  C.P. 8888 -- Succ. Centre-Ville, Montr{\'e}al H3C 3P8 (Canada) \\
%  ~\\
  \{Baptiste.Ducatel, KaytoueM, MarcuolF, Napoli\}@loria.fr, Szathmary.L@gmail.com
}
\maketitle
%%%%  Pas de num\'erotation sur la page de titre
\thispagestyle{empty}
\subsection*{R\'esum\'e}
{\em
Conçu à l'origine pour une étude de cohorte, \coron est devenu une plate-forme de fouille de données à part entière, qui incorpore une riche collection d'algorithmes pour l'extraction de motifs (fréquents, fermés, générateurs, etc.) et la génération de règles d'association à partir de données binaires, ainsi que divers outils de pré- et post- traitements.

}
\subsection*{Mots Clef}
Extraction de connaissances, fouille de données, motifs fréquents et rares, règles d'association

%\subsection*{Abstract}
%{\em
%\coron is a domain and platform independent data-mining toolkit, implementing a collection of symbolic mining methods for itemset extraction and association rule %generation. In addition, \coron provides support for preparing and filtering data, and for interpreting the extracted patterns. 
%}
%\subsection*{Keywords}
%Knowledge discovery, data mining, frequent and rare itemsets, association rules

\section{Aper\c{c}u}
Né d'un besoin logiciel pour une étude de cohorte \cite{EGC06}, \coron est maintenant une plate-forme logicielle d'extraction de connaissances à part entière, utilisée dans divers domaines, voir par exemple \cite{Daquin07,Kaytoue08,Ignatov08}. Destinée à un usage scientifique et pédagogique, la plate-forme \coron s'articule autour de plusieurs modules pour la préparation puis la fouille de données, le filtrage et l'interprétation des unités extraites. Ainsi, à partir de données binaires (possiblement issues d'une discrétisation), \coron permet d'extraire des motifs (fréquents, fermés, etc.) puis de générer des règles d'association (non redondantes, informatives, etc.). Le système englobe ainsi des algorithmes classiques mais aussi spécifiques et propres à la plate-forme \cite{DS08,IDA09}. \coron est librement disponible au t\'{e}l\'{e}chargement \`{a} \url{http://coron.loria.fr}. Essentiellement programm\'{e} en Java 6.0 et r\'{e}dig\'{e} en anglais, il est compatible avec Unix/Mac/Windows et s'utilise en ligne de commande.

\section{Architecture}
\begin{figure}[b!] \centering
\includegraphics[width=\columnwidth]{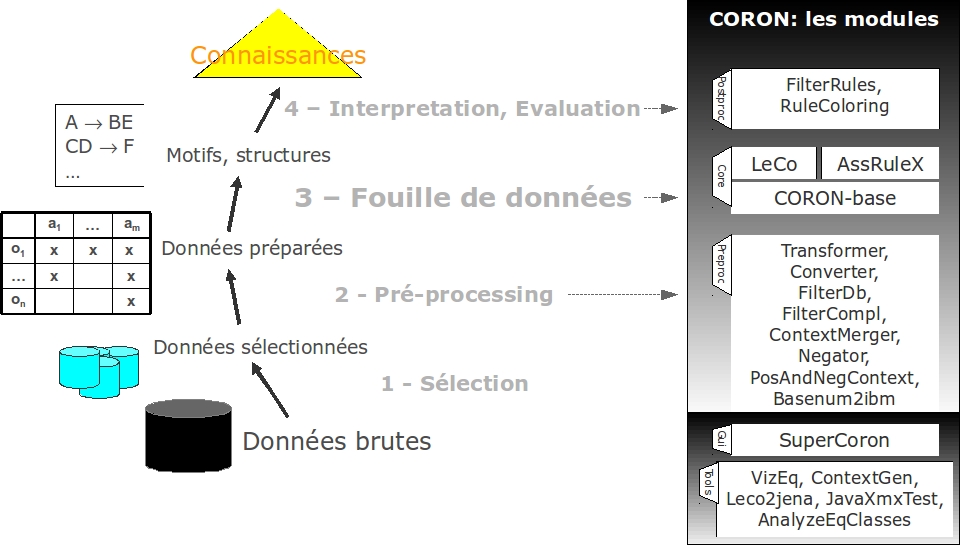} 
\caption{Architecture de la plate-forme \coron, en relation avec les \'{e}tapes du processus d'extraction de connaissances.}
\label{loop}
\end{figure}

\coron est structur\'{e} en divers modules d\'{e}di\'{e}s \`{a} chaque \'{e}tape du processus d'extraction de connaissances (Fig. \ref{loop}).

\paragraph{Modules de pré-traitement.}
Ces modules offrent de nombreux outils de formatage et de manipulation des donn\'{e}es brutes. Les données sont décrites par des tables binaires matérialisées sous forme tabulaire dans des fichiers textes bruts : des individus en lignes possèdent ou non des propriétés en colonnes. Les opérations possibles sont principalement: (i) la discrétisation de données numériques, (ii) la conversion de format de fichiers, (iii) la création du complément et du transposé d'une table binaire, ou encore (iv) diverses opérations de projection de la table.

\paragraph{Modules de fouille de donn\'{e}es.}
Découvrir des motifs ou des règles d'association est une tâche très populaire en fouille de données et plus généralement en intelligence artificielle. Par exemple, $A \rightarrow BE$, accompagnée de mesures comme le support et la confiance, permet de refléter les conditions dans lesquelles il est licite de dire ``les individus qui ont la propriété $A$ ont également les propriétés $B$ et $E$''. Pour construire ces règles, il faut généralement d'abord extraire des motifs d'intér\^et.  Un motif reflète les conditions dans lesquelles un ensemble de propriétés appara\^it. Par exemple, le motif $ABE$ peut \^etre à l'origine de $A \rightarrow BE$.

Ainsi, les modules de fouille de données de \coron permettent respectivement
\begin{itemize}
\item l'extraction de motifs : fr\'{e}quents, fermés fréquents, rares, générateurs, etc. à l'aide d'une collection d'algorithmes de la littérature s'appuyant sur différentes stratégies de parcours de l'espace de recherche (par niveau, profondeur, hybride).
\item la génération de r\`{e}gles d'association : fr\'{e}quentes, rares, ferm\'{e}es, informatives, minimales non redondantes réduites, de la base de Duquenne-Guigues, etc. Ces règles sont associées à un ensemble de mesures comme le support, la confiance, le lift et la conviction.
\item la construction d'un treillis, structure sous-jacente de l'ensemble des motifs extraits, à partir d'une table binaire.
\end{itemize}

\paragraph{Modules de post-traitement.}
Les unités extraites dans l'étape précédente peuvent \^etre très nombreuses et cacher de ce fait des unités intéressantes. Ainsi, diverses étapes permettent de les filtrer, de préférence en interaction avec un expert du domaine des données. L'analyste peut ainsi \'{e}valuer les r\'{e}sultats en utilisant un outil de filtrage (syntaxique ou fonction de la longueur des pr\'{e}misses et cons\'{e}quents des r\`{e}gles obtenues) ou en se concentrant sur les $k$ meilleures unités extraites, en regard d'une mesure qu'il aura jug\'{e} pertinente. Un focus syntaxique est \'{e}galement possible par le biais d'un outil de colorisation des propriétés ciblées.

\paragraph{Bo\^{i}te-\`{a}-outils.} Finalement, des modules auxiliaires permettent, par exemple, la visualisation de classes d'\'{e}quivalence, la g\'{e}n\'{e}ration al\'{e}atoire de jeux de donn\'{e}es ou l'optimisation de la m\'{e}moire \`{a} solliciter pour le fonctionnement du programme.

\section{\coron en pratique}
\paragraph{Application aux \'{e}tudes de cohorte.}Le suivi de la cohorte ``STANISLAS'' a \'{e}t\'{e} conduit \`{a} Nancy, conjointement par une \'{e}quipe de m\'{e}decins (INSERM) et par des membres associ\'{e}s \`{a} la conception du syst\`{e}me \coron. L'objectif de l'analyse \'{e}tait de caract\'{e}riser le profil g\'{e}n\'{e}tique associ\'{e} au syndrome m\'{e}tabolique, un trouble regroupant des facteurs de risque pr\'{e}disposant aux maladies cardiovasculaires et au diab\`{e}te de type II. L'utilisation de la plate-forme \coron a permis de faire \'{e}merger un profil in\'{e}dit : une personne poss\'{e}dant l'all\`{e}le rare pour le polymorphisme APOB71Thr/Ile serait plus fr\'{e}quemment atteinte par le syndrome m\'{e}tabolique \cite{EGC06}. La m\'{e}thodologie mise en place autour de cette premi\`{e}re exp\'{e}rience est aujourd'hui affin\'{e}e dans le cadre d'une seconde \'{e}tude de cohorte. L'objectif est ici d'\'{e}valuer la valeur pr\'{e}dictive d'un acide amin\'{e}, l'homocyst\'{e}ine, dans l'apparition de maladies li\'{e}es au vieillissement. Les donn\'{e}es sont recueillies aupr\`{e}s d'une population rurale m\'{e}diterran\'{e}enne, la cohorte OASI. 

\paragraph{Autres applications.}
\coron est utilisé pour des tâches comme l'extraction de connaissances d'adaptation en raisonnement à partir de cas \cite{Daquin07}, l'étude de données d'expression de gènes \cite{Kaytoue08}, la comparaison de méthodes de construction de treillis de concepts à partir de données numériques avec et sans binarisation \cite{Kaytoue09}, la classification dynamique pour la recherche d'information sur le web \cite{Coria08}, la recommandation de publicité sur internet \cite{Ignatov08},  l'intégration de donnés biologiques \cite{Coulet08} et bien sûr l'étude de cohortes \cite{EGC06}.

\section{Travaux en cours}
Les travaux en cours concernent principalement l'intégration de \coron dans la plateforme de fouille Knime \cite{knime}, dont la popularité est croissante. Ainsi, \coron bénéficiera de nombreux avantages (voir \url{http://www.knime.org}). L'intégration de méthodes d'extraction d'unités à partir de données complexes, sans discrétisation comme dans \cite{Kaytoue09} pour les données numériques, est à l'étude. Enfin, un forum est mis en place pour recueillir lers retours d'exp\'{e}riences des utilisateurs de \coron (\url{http://coron.loria.fr/forum/}).

\end{document}